\documentclass[aps,prl,twocolumn,showpacs]{revtex4}
\usepackage{epsfig}
\usepackage{graphicx}

\begin{document}

\DeclareGraphicsExtensions{.eps,.EPS}

\title{Anisotropic excitation spectrum of a dipolar quantum Bose gas}
\author{G. Bismut, B. Laburthe-Tolra, E. Mar\'echal, P. Pedri, O. Gorceix and L. Vernac}
\affiliation{Laboratoire de Physique des Lasers, UMR 7538 CNRS,
Universit\'e Paris 13, 99 Avenue J.-B. Cl\'ement, 93430
Villetaneuse, France}

\begin{abstract}
We measure the excitation spectrum of a dipolar Chromium Bose Einstein Condensate with Raman-Bragg spectroscopy.  The energy spectrum depends on the orientation of the dipoles with respect to the excitation momentum, demonstrating an anisotropy which originates from the dipole-dipole interactions between the atoms. We compare our results with the Bogoliubov theory based on the local density approximation, and, at large excitation wavelengths, with numerical simulations of the time dependent Gross–Pitaevskii equation. Our results show an anisotropy of the speed of sound.

\end{abstract}

\date{\today}
\maketitle

Interactions play a major role in the physics of Bose Einstein Condensates (BECs) made of trapped neutral atoms. Attractive interactions may lead to collapse \cite{Burnett}, while repulsive interactions confer a collective nature to excitations \cite{collectiveKetterle}, and lead to superfluidity \cite{SuperfluidflowKetterle}. In the first produced BECs only contact interactions played a significant role. The study of long-range anisotropic (partially attractive) dipole dipole interactions (DDIs) in quantum gases was opened by the production of Chromium BECs \cite{CrBEC}, followed more recently by experiments with Dysprosium \cite{DysprosiumBEC,DysprosiumFermiSea} and Erbium \cite{Erbium}. DDIs introduce an anisotropy in the expansion dynamics \cite{CrBECexpansion} and the collective excitations of a Cr BEC \cite{Bismut}. Moreover when DDIs overwhelm contact interactions a BEC undergoes a characteristic $d$-wave like implosion which reveals the structure of DDIs \cite{CrBECcollapse,Erbium}. In addition, anisotropic dipolar interactions may lead to the study of anisotropic superfluidity in Bose \cite{Bohn} or Fermi gases \cite{dipolarFermion}, with possible analogies to anisotropic superconductivity in cuprates. In this paper we study elementary excitations of a Cr BEC and show an anisotropic excitation spectrum,  providing a signature of an anisotropy of the speed of sound and therefore, according to the Landau criterion, a possibility for anisotropic superfluidity.

\begin{figure}[h]
\centering
\includegraphics[width=3.2in]{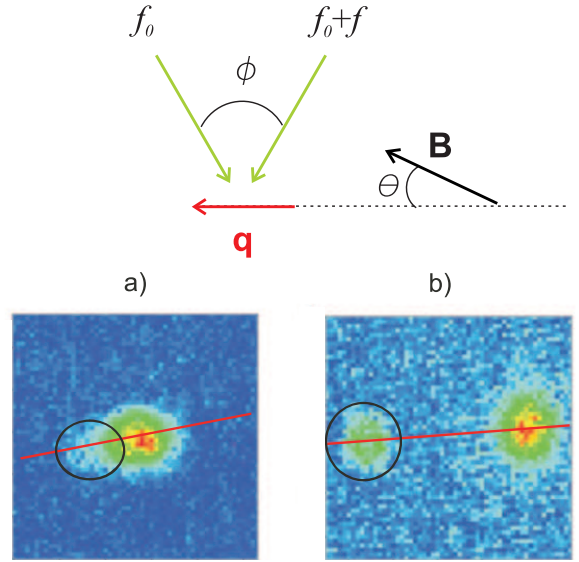}
\caption{\setlength{\baselineskip}{6pt} {\protect\scriptsize (Color online) Principle of the experiment and absorption pictures after a Bragg pulse. The two Bragg beams are coupled to the BEC with an angle $\phi$, and transfer a momentum $\hbar$\textbf{q} and an energy $hf$ to the excited fraction. Due to DDIs, the excitation spectrum depends on the angle $\theta$ between \textbf{q} and \textbf{B}. The two absorption pictures show in false color the density after Bragg pulse and time of flight, and solid lines indicate the direction of momentum transfer. In a) the small value of $\phi$ ($14\deg$), and hence of $q$, barely allows spatial separation of the excited fraction, contrary to the case of large $\phi$ ($82\deg$) in b).}} \label{Bragg}
\end{figure}

For investigating elementary excitations with a well defined momentum, Raman-Bragg spectroscopy has developed into a powerful instrument \cite{Ozeri}. Following the first seminal demonstration that rapidly followed the creation of atomic BECs \cite{BraggKetterle1,BraggKetterle2}, more systematic series of experiments gave a full picture of the BEC excitation spectrum from the low-energy phonon-like excitations to the high-energy single-particle regime \cite{Steinhaeuer}. Despite several reviews and theoretical papers pointing out the interest of the excitation spectrum of bosonic gases featuring DDIs \cite{Lahaye,Lima,Muruganandam}, experimental results are not yet available \cite{noteEsslinger}. Chromium atoms away from a Feshbach resonance are particularly suitable to study bulk properties of dipolar BECs since DDIs are non negligible without leading to collapse. In this work, we provide the first evidence that long-range dipolar interactions induce an anisotropy of a BEC excitation spectrum using Raman-Bragg spectroscopy. The spectrum is anisotropic as the resonance condition depends on the angle, $\theta$, between the polarization axis, parallel to the magnetic field $\textbf{B}$, and the excitation wave-vector $\textbf{q}$ (see Fig \ref{Bragg}). We probe this anisotropy throughout the whole excitation spectrum with differential measurements for two orthogonal orientations of the magnetic field, either parallel or orthogonal to $\textbf{q}$.

The anisotropy of the excitation spectrum for an homogenous dipolar BEC is well understood within the Bogoliubov theory \cite{Lahaye}:

\begin{equation}
 \epsilon(\mathbf{q})= \left[ \frac{\hbar^2 q^2}{2m} \left(\frac{\hbar^2 q^2}{2m} + 2g n \left(1+\epsilon_{dd}\left(3\cos^2\theta-1\right) \right)\right)\right]^{1/2}
\label{dispdd}
\end{equation}
where $\epsilon(\mathbf{q})$ and $\textbf{q}$ are respectively the energy and the wave vector of the excitation, $m$ the atom mass, $n$  the atomic density, and $g = 4 \pi \hbar^2 a /m $ (with $a$ the scattering length); $\epsilon_{dd}$ is the dimensionless parameter scaling the relative importance of dipolar interactions with respect to contact interactions. For magnetic DDIs,   $\epsilon_{dd}= \mu_0 \mu_m^2 m / 12   \hbar^2 a$ ($\mu_0$ is the vacuum permeability, $\mu_m$ the atom magnetic moment equal to 6 Bohr magnetons for Cr). In absence of DDIs ($\epsilon_{dd}=0$), one recovers the known spectrum for a BEC with only contact interactions \cite{Stringari}. For Cr  $\epsilon_{dd}$ is 0.16 (with $a=102.5a_0$ \cite{ControlPRA}), which allows to explore the anisotropic character of eq. (\ref{dispdd}) more easily than for alkali BECs (e.g. $\epsilon_{dd}=0.01$ for Rb). The angular dependance of eq. (\ref{dispdd}) can be interpreted as an attractive contribution (resp. repulsive) of DDIs to the excitation energy for the perpendicular (parallel) case \cite{Lahaye}. It is a direct consequence of the momentum sensitivity of DDIs, in contrast to contact interactions which are momentum-insensitive.

According to eq.(\ref{dispdd}), the sound velocity defined as $c_{\theta}= \lim_{q\rightarrow 0} \frac{\epsilon(\mathbf{q})}{q}$, becomes anisotropic. It is maximal in the parallel ($\theta=0$) geometry, while it is minimal in the perpendicular ($\theta=\pi/2$) geometry. Defining the velocity without DDIs $c_0=(gn/m)^{1/2}$, one obtains $c_{\parallel} =c_0 (1+2 \epsilon_{dd})^{1/2}$, and $c_{\perp} =c_0 (1- \epsilon_{dd})^{1/2}$, so that $c_{\parallel}/c_{\perp}= 1.25$ for Cr. This dipolar effect is larger than previous observations\cite{CrBECexpansion,Bismut}, where effects are in the few $\%$ range, much less than $\epsilon_{dd}$, as a result of an angular averaging of DDIs in the BEC \cite{CrBECexpansion,Santos-Pfau}. The situation is quite different for the BEC excitations described by eq. (\ref{dispdd}), as there is no such angular averaging: $\theta$ has the same value for all the atoms.

\begin{figure}[h]
\centering
\includegraphics[width=3.4in]{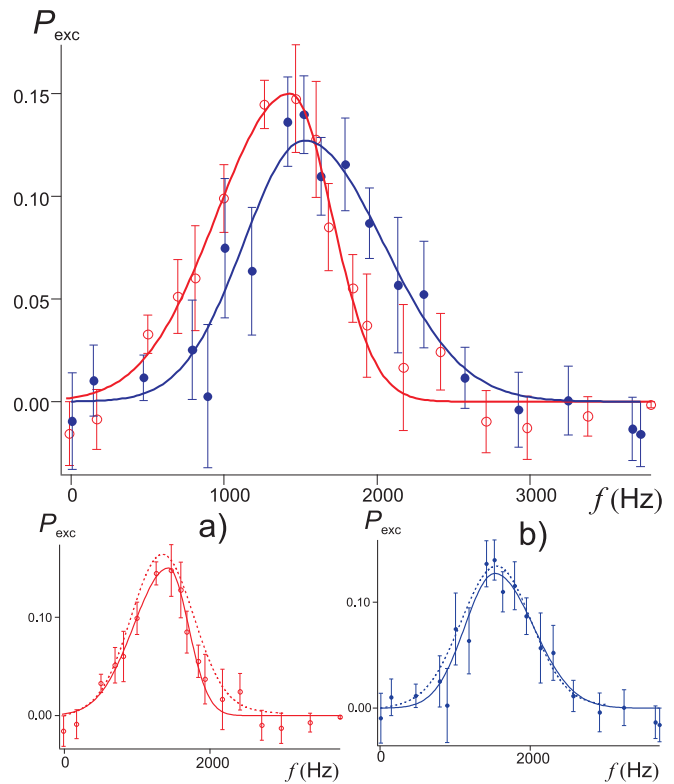}
\caption{\setlength{\baselineskip}{6pt} {\protect\scriptsize (Color online) Excitation spectra for $q\xi_0=0.8$ ($\phi=14\deg$). The excited fraction $P_{\rm exc}$ is plotted versus the detuning frequency $f$ between the two Bragg beams, for $\theta=0$ (filled circles) and $\theta=\pi/2$ (open circles). The two solid lines are asymetric gaussian fits to the data (see text). The error bars represent the 1 sigma statistical uncertainty associated to three measurements. The two spectra are shown separately below ( a): $\theta=\pi/2$, b): $\theta=0$) along with the results of our linear response theory (dashed lines) which has no adjustable parameters.}} \label{Spectre}
\end{figure}

To obtain the experimental excitation spectra, we first create a chromium BEC in a crossed-beam dipole trap. Details on the experimental setup and the procedures can be found in \cite{CrBECus}. The BEC, comprising about 10000 atoms polarized in the absolute ground (Zeeman) state, is confined at the bottom of the trap with frequencies  ($\omega_x$, $\omega_y$, $\omega_z$) = $2 \pi \times (145, 180, 260)$ Hz, within a magnetic field of B=50 mGauss. To impart a momentum $q$ to a fraction of the condensed atoms, we use two intersecting phase-locked focused laser beams derived from the same single-mode solid-state laser ($\lambda=532$ nm). Two acousto-optic modulators (AOMs) shift the beam frequencies by respectively $\Omega$ and $\Omega+2\pi f$, with $\Omega= 2 \pi \times 80$ MHz. At the atom location, their respective waists are 40 and 32 $\mu$m while their powers are between 500 and 1000 $\mu$W. The beam intensities are chosen to limit the excited fraction to about 15 $\%$  which ensures a good trade-off between the validity of the perturbative Bogoliubov approach for theoretical interpretation  and the signal to noise ratio. The angle $\phi$ between the beams propagating directions sets the momentum value $q$ ($q=2\pi/\lambda\times \sin(\phi/2)$), while the energy difference $h f$, equal to the energy of the excitation, is adjustable at will. Although optical access restricts the accessible values for $\phi$, thus also that for $q$, we have been able to probe the dispersion relation from the phonon to the free particle regime.

The two beams hereafter referred to as the Raman beams are switched-on only for a brief pulse-time. Raman transitions are efficiently driven between the rest ground state and excited states with momentum $q$ and energy $hf$ when the beam photon energy difference matches the excitation energy $i.e.$ when the Raman-Stokes resonance condition is met. Since the process can be interpreted as diffraction of the matter-wave onto the moving lattice created by the Raman-beams, the whole process is often referred to as Bragg spectroscopy of the excitation spectrum of the BEC. To ensure that the momentum of the atoms is solely set by the Bragg pulse (and is not modified by trap dynamics), the pulse duration $\tau_{\rm Bragg}$ should be far less than the oscillation periods in the trap \cite{BraggKetterle1}. Meeting this requirement causes a strong Fourier broadening and we found a trade-off by setting $\tau_{\rm Bragg}=1.5$ ms: the Fourier broadening (half width at 1/$e^2$=300 Hz) is then smaller that the experimental spectra widths, and spatial separations between the excited fraction and the ground state are close to the ones expected for the Bragg momentum transfer (see Fig \ref{Bragg}).

We measure the atom momentum distribution by releasing the atoms from the trap after the Bragg pulse and by performing absorption imaging after an expansion time of 5 ms (see Fig \ref{Bragg}). From the density profiles of these distributions along the excitation direction, we infer the excited fraction $P_{\rm exc}=[N_{\rm exc}/(N_{\rm exc}+N_0 )]$ where $N_{\rm exc}$ is the number of excited atoms and $N_0$ the number of atoms remaining in the ground $q\sim0$ state. We record Bragg spectra by monitoring the excited fraction versus $f$ for a given value of $\phi$. It is clear from Fig \ref{Spectre} that the excitation energy spectrum for parallel polarization is shifted towards high frequency with respect to the spectrum for orthogonal polarization, as expected from eq. (\ref{dispdd}).

The results of eq. (\ref{dispdd}) hold only in the homogeneous case. With trapped gases, two kinds of finite size effects have to be taken into account, which are sources of spectral broadenings: the density inhomogeneity in the trap, and the non zero-width momentum distribution (inducing Doppler effect), which are dominant at respectively low and high excitation momentum \cite{Zambelli}. To account for our experimental results, we extended the theory developed in \cite{Zambelli} based on the local density approximation (LDA), by including DDIs. The validity of LDA with DDIs, discussed in \cite{Lima}, is guaranteed if $q\gg1/R_{TFmin}$ \cite{BismutThese}, with $R_{TFmin}$ the smallest BEC Thomas Fermi (TF) radius. Then the BEC can be considered as a locally homogeneous 3D system \cite{note basses D}.  $q\gg1/R_{TFmin}$ also ensures that effects of energy discretization can be safely neglected.

To compare with our experimental spectra (see Fig \ref{Spectre}), we take into account the Doppler width and make a convolution between the corresponding Gaussian and the LDA excitation spectrum \cite{SKurn Revue}. The Fourier broadening due to the finite excitation duration is also taken into account within the linear response theory \cite{Brunello}. The two photon Raman frequency which sets the excitation amplitude \cite{Steinhaeuer} is estimated from a calibration of the lattice depth created by the Raman beams \cite{noteCalibration}. The agreement with theory (with no adjustable parameter) for the spectra line shapes is good except for the two lowest experimental values of $q$. For the excitation amplitude, the discrepancy remains at most in the $50\%$ range for all spectra. All features are in very good agreement in the case of Fig \ref{Spectre}.

\begin{figure}[h]
\centering
\includegraphics[width=4in]{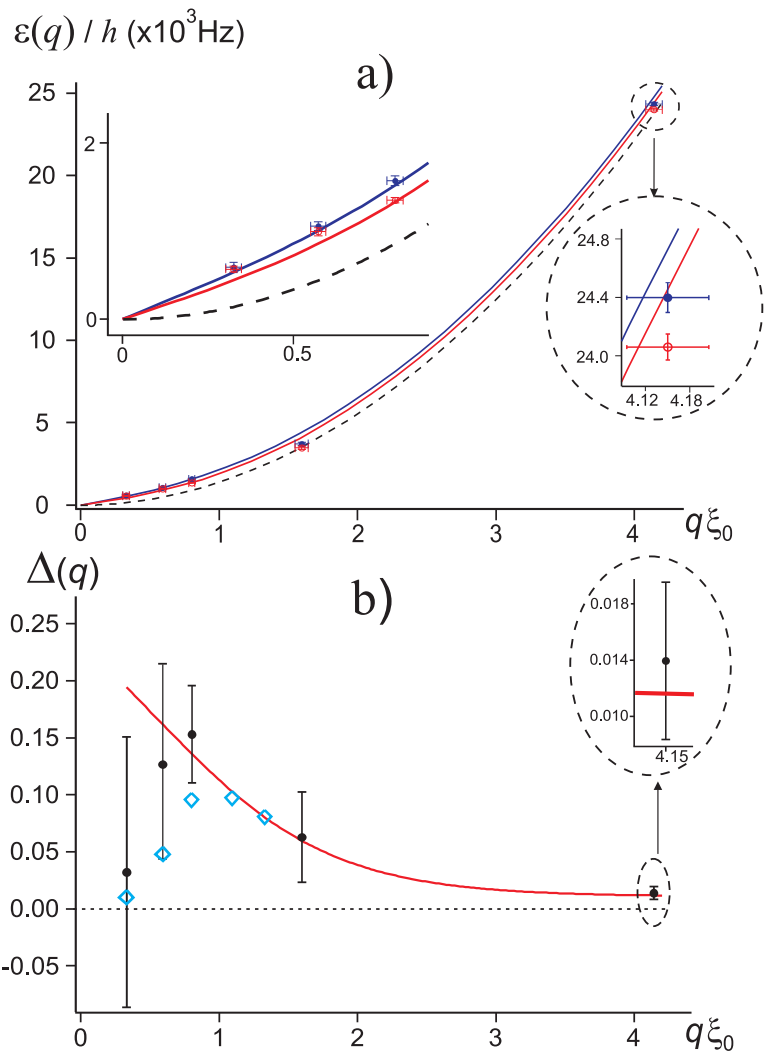}
\caption{\setlength{\baselineskip}{6pt} {\protect\scriptsize (Color online) a) Energy spectra: the energy $\epsilon(\mathbf{q})$ divided by $h$ is plotted as a function of Bragg-transferred momentum $q$. Circles correspond to the experimental data: for a given value of $q$, the top (resp. bottom) circle corresponds to the parallel (perpendicular) case. The point at $q\xi_0=1.6$ corresponds to a four photon process (see text). The two solid lines are the results of LDA calculations for resp. parallel (top line) and orthogonal (bottom line) case. The dashed line gives the results for a free particle. Inset: zoom for low $q$ values. b) Relative shift $\Delta(q)$ of the excitation energies (see text) as a function of $q$. Black points: results of fits of the experimental data, with error bars. Solid line: results of LDA calculations. The diamonds show results of numerical simulations. For a) and b), vertical error bars represent 1$\sigma$ statistical uncertainty, while horizontal ones correspond to uncertainties on values of $\phi$.}} \label{DeltaEqDisp}
\end{figure}

We have performed a study of effects of DDIs by probing the excitation spectrum as a function of $q$ from the phonon regime to the free particle regime (see Fig \ref{DeltaEqDisp}). We define the experimental value of $\epsilon(\mathbf{q})$ as the mean value of the excitation spectra, obtained by using asymmetric Gaussian fits; this procedure empirically takes into account the non-symmetric nature of the excitation spectra which is expected from the Thomas Fermi distribution (see for example \cite{Zambelli}). We define $\epsilon_{\parallel}(q)$ (respectively $\epsilon_{\perp}(q)$) the mean value obtained for the parallel (orthogonal) case. Fig \ref{DeltaEqDisp}a) shows the corresponding data. We use the DDI-independent healing length $\xi_0=\hbar/(2mgn_0)^{1/2}$ (with $n_0$ the measured BEC peak density) to normalize the $q$ axis: $\xi_0$ indicates the frontier between the phonon domain ($q\xi_0< 1$) and the single-particle domain ($q\xi_0\gg1$). The data correspond to four $\phi$ angles. To fill the gap between the accessible low ($6^{\circ}$, $10^{\circ}$ and $14^{\circ}$) and the high ($82^{\circ}$) values of $\phi$, we realized a four-photon Raman-like excitation \cite{Bill} to obtain data for another $q$ value ($q=4\pi/\lambda\times \sin(14^{\circ}/2)$, and $q\xi_0=1.6$). The horizontal error bars are mainly set by the uncertainty on the experimental value of $\phi$, and the vertical ones are determined by applying a bootstrapping procedure \cite{bootstrapping} to the experimental data. We compare predictions of our theory assuming a constant ($10^4$) number of atoms with the corresponding experimental data in Fig \ref{DeltaEqDisp}a) \cite{noteTheory}. We obtain a good general agreement between data and theory with no free parameter; some of the discrepancies may originate from an underestimation of the uncertainties on the values of $\phi$.

To emphasize how the two dispersion curves differ, we plot in Fig \ref{DeltaEqDisp}b) the relative variation  of $\epsilon(\mathbf{q})$, $\Delta(q)=2\frac{\epsilon_{\parallel}(q)-\epsilon_{\perp}(q)}{/\epsilon_{\parallel}(q)+\epsilon_{\perp}(q)}$. We measure an anisotropy of the excitation spectrum of the Cr-BEC, both in the low $q$ limit, and the high $q$ limit. As systematic effects on $\Delta(q)$ related to changes in trapping frequencies and atom number induced by the change in $\theta$ remain negligible (from at most $1.5\%$ at low $q$ to below $0.1\%$ at high $q$), our results demonstrate a DDI induced anisotropy of the excitation spectrum. The agreement with our linear response theory is relatively good, except for the lowest value of $q$. This is not surprising since then the excitation wavelength is larger than the BEC Thomas Fermi diameter (equal to $7$ $\mu$m). For long wavelengths, discrete modes can be excited, for which the effects of DDIs are much smaller than for phonons, as observed in \cite{Bismut}. We have therefore solved numerically the time dependent 3D Gross-Pitaevskii equation including contact interactions and DDIs. Our numerical simulations are in rather good agreement with our experimental data (see Fig \ref{DeltaEqDisp}b)), and show that relative-shifts significantly smaller than the ones obtained with LDA theory are expected as the excitation wavelength increases.

We finally rely on LDA theory to extract quantitative values for the sound velocity from experimental data. In our case the extensive experimental study of the linear phonon part of the spectrum is not accessible due to the small size of our BEC (contrary to \cite{Steinhaeuer}). We therefore chose to derive sound velocities from the data at $q\xi_0=0.8$, as this is the lowest value of $q$ for which we find a good agreement with LDA theory. The corresponding excitation frequency (around 1.5 kHz) is significantly
higher than the trap frequencies, which explains why energy discretization effects are
small. From the experimental values $\epsilon_{\parallel}/h=(1.57\pm0.05)$ kHz and  $\epsilon_{\perp}/h=(1.35\pm0.03)$ kHz, we derive sound velocities through the linear response theory. The Feynman law (see \cite{Steinhaeuer}) relates the energy $\epsilon_{\mathbf{q}}$ to the
(LDA averaged) static structure factor $S_{\mathbf{q}}$: $\epsilon_{\mathbf{q}}=\frac{\epsilon_{\mathbf{q}}^0}{S_{\mathbf{q}}}$,
with $\epsilon_{\mathbf{q}}^0=\hbar^2q^2/2m$ the free particle energy; and $S_{\mathbf{q}}$ is related to $c_{\theta}$ through:
$S_{\mathbf{q}}=\frac{15}{4}\left\{\frac{3+\alpha_{\mathbf{q}}}{4{\alpha_{\mathbf{q}}}^2}-\frac{(3+2\alpha_{\mathbf{q}}-{\alpha_{\mathbf{q}}}^2)}{16{\alpha_{\mathbf{q}}}^{5/2}}\left[\pi+2\arctan\left(\frac{\alpha_{\mathbf{q}}-1}{2\sqrt{\alpha_{\mathbf{q}}}}\right)\right]\right\}$
where
$\alpha_{\mathbf{q}}=2 m (15\pi c_{\theta}/32)^2/\epsilon_{\mathbf{q}}^0$.
We obtain $c_{\perp}= 2.06\pm0.05$ mm.s$^{-1}$ and $c_{\parallel} = 2.64\pm0.1$ mm.s$^{-1}$, in good agreement with LDA predictions in our experimental conditions equal to $2.02\pm0.05$ mm.s$^{-1}$ and $2.53\pm0.05$ mm.s$^{-1}$ respectively ($c_{\theta,LDA}=\frac{32}{15\pi}\left[\frac{gn_0}{m}(1+\epsilon_{dd}(3\cos^2\theta-1))\right]^{1/2}$).

In this article, we have shown that the excitation spectrum of a spin-polarized chromium BEC is anisotropic as a consequence of the dipolar character of the interactions between the atoms. As long as the wave vector is not too small compared to the inverse of the BEC size, our experimental results are correctly accounted for by a linear response theory based on the Bogoliubov approach. The demonstrated existence of an anisotropic speed of sound raises the question of anisotropic superfluidity, for example how excitations in different directions couple. Extension of our work to BECs with larger DDIs could allow to discover rotonic features in the excitation spectrum \cite{Rotons} of quasi 2D BECs.

\vspace{0.5cm}

Acknowledgements: LPL is Unit\'e Mixte (UMR 7538) of CNRS and
of Universit\'e Paris 13. We acknowledge financial support from Conseil R\'{e}%
gional d'Ile-de-France (CPER), Minist\`{e}re de
l'Enseignement Sup\'{e}rieur et de la Recherche and
IFRAF.


\begin{thebibliography}{99}
\bibitem{note basses D} In lower dimensions systems, effects of DDIs on the Bogoliubov excitation energy are totally different than in the 3D case (\cite{Muruganandam}).
\bibitem{noteTheory} As proven in \cite{Zambelli}, Doppler broadening does not change the average excitation energy predicted by LDA, so that calculations in Fig \ref{DeltaEqDisp} are also in fact LDA results.
\end{thebibliography}
\end{document}